\documentstyle[psfig,12pt,twoside,fleqn,espcrc1]{article}

\newcommand{\AmS}{{\protect\the\textfont2
  A\kern-.1667em\lower.5ex\hbox{M}\kern-.125emS}}
\newcommand{\sla}{\!\!\!\!/ \,}
\title{Gluon condensate, quark propagation, and dilepton production in the 
quark-gluon plasma}

\author{M.G. Mustafa\address{Institut f\"ur Theoretische Physik,
Universit\"at Giessen, 35392 Giessen, Germany}\thanks{Humboldt Fellow} 
A. Sch\"afer\address{Institut f\"ur Theoretische Physik, 
Universit\"at Regensburg, 93040 Regensburg, Germany}, and
M. H. Thoma\address{ECT*, Villa Tambosi, Strada delle Tabarelle 286, 
38050 Villazzano (Trento), Italy}\thanks{Heisenberg Fellow}}

\begin{document}
\maketitle

\begin{abstract}
A calculation of the thermal quark propagator 
is presented taking the gluon condensate above the critical 
temperature into account. The quark dispersion relation 
and the dilepton production following from this propagator are derived.
\end{abstract}

\vspace*{1cm}

As an alternative method to lattice and perturbative QCD 
we suggest to include the gluon condensate into the parton propagators
\cite{ref4a}. 
In this way non-perturbative effects are taken into account within 
the Green functions technique. 

In the case of a pure gluon gas with energy density $\epsilon $ and 
pressure $p$ the gluon condensate can be related to the interaction 
measure $\Delta =(\epsilon -3p)/T^4$ via \cite{ref3}
\begin{equation}
\langle G^2 \rangle _T=\langle G^2  \rangle _0-\Delta T^4,
\label{e5}
\end{equation}
where $G^2\equiv (11\alpha _s/8\pi) : G^a_{\mu \nu}G_a^{\mu \nu}:$ 
and $\langle G^2  \rangle _0 = (2.5 \pm 1.0)\> T_c^4$ is the zero 
temperature condensate. Here $G_{\mu \nu}^a$ is the field strength tensor and 
$T_c$ the critical temperature of the phase transition to 
the quark-gluon plasma (QGP).

At zero temperature the quark propagator containing the gluon condensate 
has been constructed already \cite{ref5}. Here we will extend these 
calculations to finite temperature.

\begin{figure}[t]
\centerline{\psfig{figure=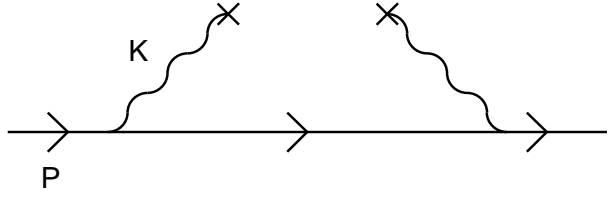,width=8cm}}
\vspace*{-1cm}
\caption{Quark self energy containing a gluon condensate.}
\end{figure} 


The full quark propagator in the QGP can be 
written by decomposing it according to its helicity eigenstates$\>$\cite{ref8}
($P=(p_0,{\vec p}\,)$, $p=|{\vec p}\, |$)
\begin{equation}
S(P)=\frac{\gamma _0-\hat p\cdot \vec \gamma}
{2D_+(P)} + \frac{\gamma _0+\hat p\cdot \vec \gamma} {2D_-(P)},
\label{e12}
\end{equation}
where 
\begin{equation}
D_\pm (P)=(-p_0\pm p)\> (1+a) - b
\label{e12a}
\end{equation}
and
\begin{eqnarray}
a & = & \frac{1}{4p^2}\> \left [tr\, (P\sla \Sigma ) - p_0\> tr\, 
(\gamma _0 \Sigma )\right ],\nonumber \\
b & = & \frac{1}{4p^2}\> \left [P^2\> tr\, (\gamma _0 \Sigma ) - p_0\> 
tr\, (P\sla \Sigma )\right ].
\label{e7} 
\end{eqnarray}

Using the imaginary time formalism and expanding the quark propagator
in Fig.1 for small loop momenta \cite{ref5}, 
i.e. $k\ll p$ and $k_0=2\pi inT=0$, we obtain \cite{ref4a}
\begin{eqnarray}
a & = & -\frac{4}{3}\, g^2\, \frac{T}{P^6}\int \! \frac{d^3k}{(2\pi )^3}\,
\left [\left (\frac{1}{3}p^2-\frac{5}{3}p_0^2\right )k^2\, \tilde 
D_l(0,k)+ \left (\frac{2}{5}p^2-2p_0^2\right )k^2\, 
\tilde D_t(0,k)\right ], \nonumber \\
b & = & -\frac{4}{3}\, g^2\, \frac{T}{P^6}\int \! \frac{d^3k}{(2\pi )^3}\,
\left [\frac{8}{3}p_0^2\, k^2\, \tilde D_l(0,k)+ \frac{16}{15}p^2\, k^2\, 
\tilde D_t(0,k)\right ],
\label{e9} 
\end{eqnarray}
where $\tilde D_{l,t}$ are the longitudinal and transverse parts of the
non-perturbative gluon propagator at finite temperature in Fig.1.   

The moments of the longitudinal and transverse gluon propagator in (\ref{e9}) 
are related to the chromoelectric and chromomagnetic condensates
via
\begin{eqnarray}
\langle {\bf E}^2\rangle _T & = & \langle :G_{0i}^aG_{0i}^a:\rangle _T
= 8T\> \int \frac{d^3k}{(2\pi )^3}\> k^2\> \tilde D_l(0,k),\nonumber \\
\langle {\bf B}^2\rangle _T & = & \frac{1}{2}\> 
\langle :G_{ij}^aG_{ij}^a:\rangle _T
= -16T\> \int \frac{d^3k}{(2\pi )^3}\> k^2\> \tilde D_t(0,k).
\label{e10} 
\end{eqnarray}
These condensates can be extracted from the expectation values of the
space- and timelike plaquettes $\Delta _{\sigma ,\tau}$ computed
on the lattice \cite{ref4}, using
\begin{eqnarray}
\frac{\alpha _s}{\pi }\> \langle {\bf E}^2 \rangle _T & = & \frac{4}{11}\>
\Delta _\tau\> T^4 - \frac{2}{11}\> \langle G^2\rangle _0,\nonumber \\ 
\frac{\alpha _s}{\pi }\> \langle {\bf B}^2 \rangle _T & = & -\frac{4}{11}\>
\Delta _\sigma\> T^4 + \frac{2}{11}\> \langle G^2\rangle _0.
\label{e11}
\end{eqnarray}

The quark dispersion relation \cite{ref8}, describing collective quark modes 
in the QGP in the presence of a gluon condensate, follows from $D_\pm (P)=0$. 
Using the lattice
results for the plaquette expectation values they have been 
determined numerically and are shown in Fig.2 for various temperatures. 

\begin{figure}[t]
\centerline{\psfig{figure=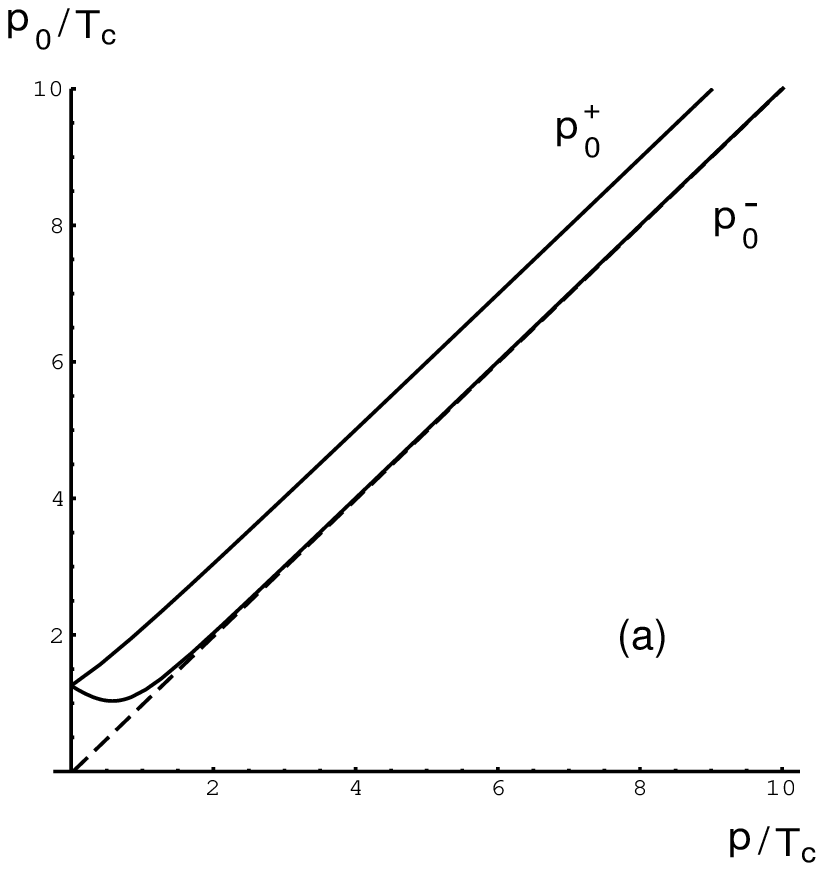,width=5.3cm}
\psfig{figure=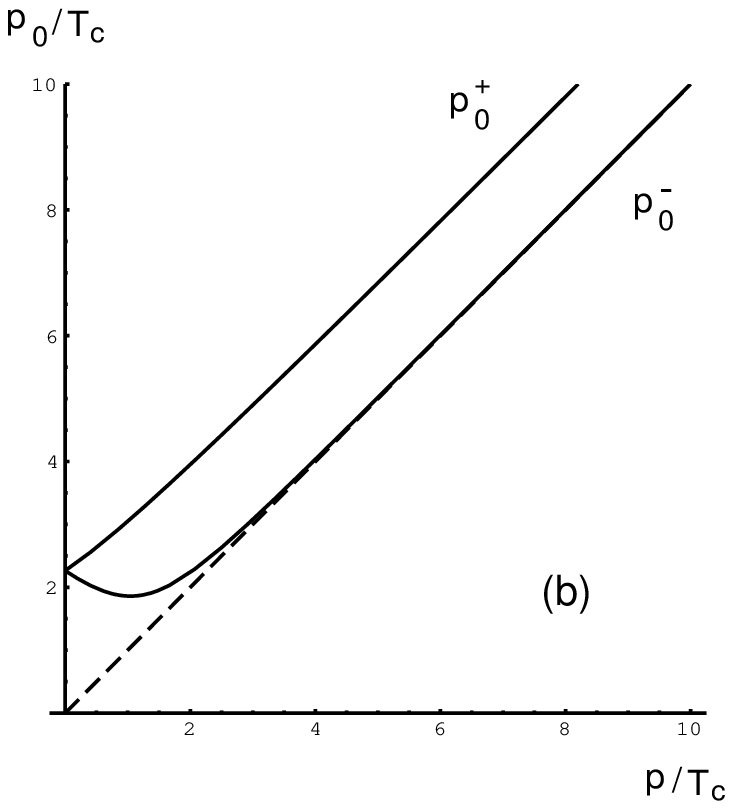,width=5.3cm}
\psfig{figure=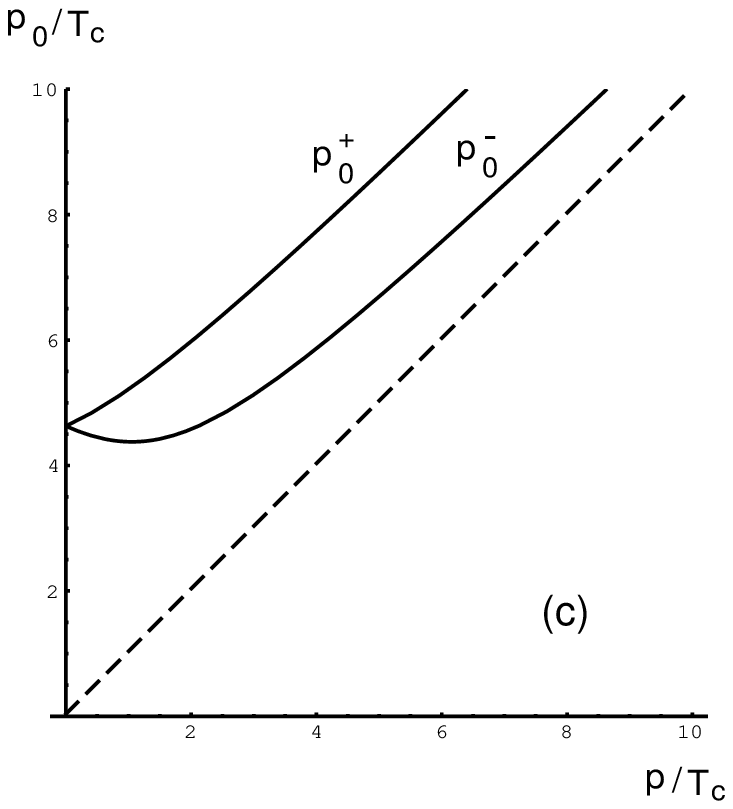,width=5.3cm}}
\vspace*{-0.5cm}
\caption{Quark dispersion relations at $T=1.1\, T_c$ (a), $T=2 \, T_c$ (b),
$T=4\, T_c$ (c) and dispersion relation of a non-interacting massless quark
(dashed lines).}
\end{figure}


The dispersions exhibit two collective quark modes. The upper branch comes 
from the solution of $D_+=0$ and the lower one from $D_-=0$. The lower
branch, showing a minimum, corresponds to a so-called plasmino, possessing 
a negative ratio of helicity to chirality, and is absent in the vacuum. 

At $p=0$ both modes start from a common effective quark mass, which is 
given by $m_{eff}=[(2\pi \alpha _s/3)\> 
(\langle {\bf E}^2\rangle _T + \langle {\bf B}^2\rangle _T )]^{1/4}$. 
In the temperature range $1.1 T_c <T< 4T_c$ we found approximately
$m_{eff}=1.15 \> T$. 

The qualitative picture of this quark dispersion relation is very similar
to the one found perturbatively in the hard thermal loop (HTL)
limit \cite{ref8}. The main 
difference is the different effective mass, which is given by 
$m_{eff}=gT/\sqrt{6}$ in the HTL approximation.

As a possible application of this effective quark propagator we 
compute the dilepton production rate from the QGP.
The dilepton production rate follows from the imaginary part of the
photon self energy in the case of two massless lepton flavors
according to \cite{ref9} 
\begin{equation}
\frac{dR}{d^4xd^4p}=\frac{1}{6\pi^4}\> \frac{\alpha }{M^2}\frac{1}{e^{E/T}-1}\>
{\rm{Im}}\Pi_\mu^\mu (P),
\label{rate}
\end{equation}
where $E=\sqrt{p^2+M^2}$ is the energy of the virtual photon with invariant
mass $M$ and momentum $p$.

Using the one-loop approximation for the photon self energy and replacing the 
bare quark propagators by the effective propagators containing the gluon 
condensate the dilepton production rate can be derived numerically 
\cite{ref10}. The result is shown in Fig.3 for $p=0$.
As in the case of soft dileptons calculated within the HTL approximation
\cite{ref8} we find peaks and gaps in the dilepton rate. The peaks 
(Van Hove singularities) are caused by the presence of the minimum in the 
plasmino dispersion. The contribution at small $M$ ending with a Van Hove
singularity comes from an electromagnetic transition from the upper branch 
to the lower branch of the quark dispersion relation. After this peak
there is a gap before the channel for plasmino annihilation 
($q_-\> \bar q_-$) opens up
with another singularity. This contribution drops quickly but at $M=2m_{eff}$
the contribution from the annihilation of collective quarks 
($q_+\> \bar q_+$) sets in.
This contribution dominates and approaches the one coming from the 
annihilation of bare quarks (Born term \cite{ref11}) at large $M$.
It should be noted that there are no smooth cut contributions 
in contrast to the HTL dilepton rate \cite{ref8}, because the 
quark self energy containing the gluon condensate has no imaginary part.

Collisions and higher order effects (bremsstrahlung) will smear
out and cover these structures to some degree. Also contributions from finite 
momenta and the space-time evolution of the fireball will wash out these
sharp structures somewhat. After all it might be interesting to 
look at low mass dileptons ($M{\buildrel < \over \sim}1$ GeV) at RHIC.
Whereas at SPS the contribution to the dilepton spectrum from the quark 
phase is suppressed by one or two orders of magnitude compared to the
hadronic contribution due to the small lifetime of the quark phase
\cite{ref12}, the quark phase is expected to dominate at RHIC. Hence these
structures coming from the presence of collective quark modes 
in the QGP might be observable and could serve as an unique signature 
for the QGP formation. 

\begin{figure}[t]
\centerline{\psfig{figure=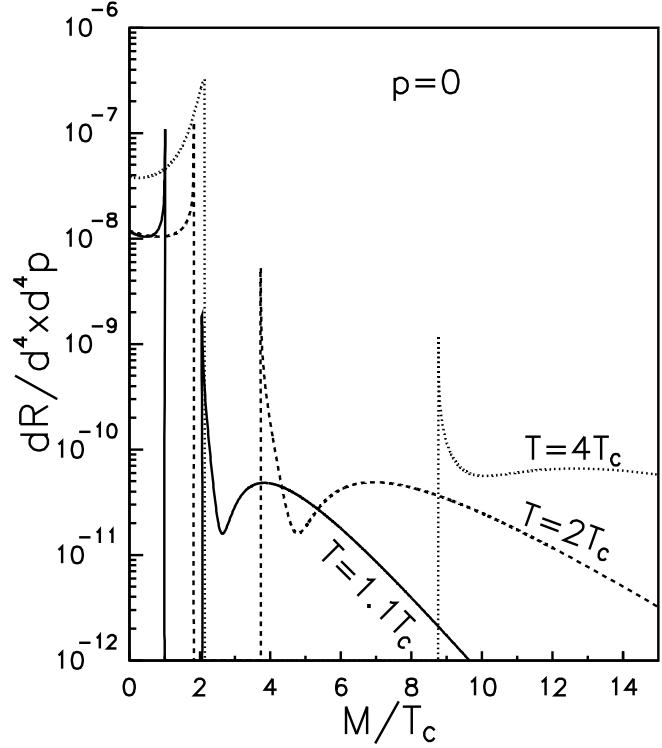,width=11cm}}
\vspace*{-3.5cm}
\caption{Dilepton production rate at zero momentum.}
\end{figure}


\end{document}